\documentclass[12pt]{iopart}
\usepackage{latexsym}
\usepackage{amssymb,amsfonts}
\usepackage{bm}        

\begin{document}

\paper{Local Hawking temperature for dynamical black holes}

\author{S. A. Hayward$^a$,
R.~Di Criscienzo$^b$, M.~Nadalini$^c$, L.~Vanzo$^c$, S.~Zerbini$^c$}

\address{$^a$Center for Astrophysics, Shanghai Normal University, 100
Guilin Road, Shanghai 200234, China} 

\address{$^b$Mc Lennan Physical Laboratories - Department of Physics,
  University of Toronto, 60 St. George Street, Toronto, ON, M5S 1A7,
  Canada}

\address{$^c$Dipartimento di Fisica, Universit\`a di Trento and INFN,
Gruppo Collegato di Trento, Italia} 
\ead{\mailto{zerbini@science.unitn.it},\mailto{vanzo@science.unitn.it}, 
\mailto{rdcris@physics.utoronto.ca},\mailto{nadalini@science.unitn.it}}

\begin{abstract}
A local Hawking temperature is derived for any future outer trapping horizon in 
spherical symmetry, using a Hamilton-Jacobi variant of the Parikh-Wilczek 
tunneling method. It is given by a dynamical surface gravity as defined 
geometrically. The operational meaning of the temperature is that Kodama 
observers just outside the horizon measure an invariantly redshifted 
temperature, diverging at the horizon itself. In static, asymptotically flat 
cases, the Hawking temperature as usually defined by the Killing vector agrees 
in standard cases, but generally differs by a relative redshift factor between 
the horizon and infinity, being the temperature measured by static observers at 
infinity. Likewise, the geometrical surface gravity reduces to the Newtonian 
surface gravity in the Newtonian limit, while the Killing definition instead 
reflects measurements at infinity. This may resolve a longstanding puzzle 
concerning the Hawking temperature for the extremal limit of the charged 
stringy black hole, namely that it is the local temperature which vanishes. In 
general, this confirms the quasi-stationary picture of black-hole evaporation 
in early stages. However, the geometrical surface gravity is generally not the 
surface gravity of a static black hole with the same parameters. 
\end{abstract}


\maketitle

\section{ Introduction}
Since the discovery of quantum black-hole radiance by Hawking 
\cite{Hawking:1974rv}, it has been  widely seen as a key area to generate and 
test ideas concerning the interface of gravity, quantum theory and 
thermodynamics. In the usual picture, a radiating black hole loses energy and 
therefore shrinks, evaporating away to an unknown fate. However, the classical 
derivations of Hawking temperature applied only to stationary black holes, 
while the above picture uses quasi-stationary arguments. In actuality, an 
evaporating black hole is non-stationary. So the question arises: is there in 
any sense a Hawking temperature for dynamical black holes? 

In seeking to generalize from statics to dynamics, one immediately faces the 
fundamental conceptual issue of what constitutes a black hole. The traditional 
definition was by an event horizon \cite{HE}. However, this is an abstract 
definition with an essentially teleological nature, unlocatable by any mortal 
observer and devoid of any local, physical relevance. Certainly it is generally 
not applicable in cosmology. Moreover, the concept leads naturally to the 
infamous information paradox. 

A more practical theory has been developed by refining the concept of apparent 
horizon to trapping horizon \cite{Hayward:1993wb,Hayward:1997jp}. Trapping 
horizons are locally defined and have physical properties such as mass, angular 
momentum and surface gravity, satisfying conservation laws \cite{AK,bhd}. They 
are a geometrically natural generalization of Killing horizons, which are 
stationary trapping horizons. In a dynamical regime, an evolving trapping 
horizon is not a null hypersurface, although it is still one of infinite 
redshift, in a sense which will be made precise below. 

For stationary black holes, Parikh \& Wilczek \cite{parikh} pioneered a 
tunneling method to derive Hawking temperature, which made precise the 
intuitive picture of Hawking radiation in terms of virtual pair creation. A 
Hamilton-Jacobi variant has recently been applied to examples of non-stationary 
black holes \cite{Visser:2001kq,NV,angh:2005,DiCriscienzo:2007fm}. The results 
all apply to trapping horizons of some sort. However, there have been 
derivations of several inequivalent temperatures or surface gravities 
\cite{Nielsen:2007ac}. Still different definitions were advocated for expanding 
cosmological black holes \cite{Faraoni:2007gq,Saida:2007ru}. 

In this paper, we apply the Hamilton-Jacobi method to general spherically 
symmetric space-times. We find that the method works if and only if there is a 
trapping horizon of the future outer type, as proposed some time ago as a local 
definition of black hole \cite{Hayward:1993wb,Hayward:1997jp}. The temperature 
so derived is given by the surface gravity as defined geometrically 
\cite{Hayward:1997jp}. We discuss the operational meaning of the temperature 
and compare with other definitions, including the usual Killing temperature in 
the static case. 

\section{Geometry}
In spherical symmetry, the area $A$ of the spheres of symmetry is a geometrical 
invariant. It is convenient to use the area radius $r=\sqrt{A/4\pi}$. A sphere 
is said to be {\em untrapped, marginal} or {\em trapped} if $g^{-1}(dr)$ is 
spatial, null or temporal respectively. If the space-time is time-orientable 
and $g^{-1}(dr)$ is future (respectively past) causal, then the sphere is said 
to be {\em future} (respectively {\em past}) trapped or marginal. A 
hypersurface foliated by marginal spheres is called a {\em trapping horizon} 
\cite{Hayward:1993wb,Hayward:1997jp}. 

The active gravitational mass $m$ \cite{MS} is defined by 
\begin{equation}
1-\frac{2m}{r}=g^{-1}(dr,dr)
\end{equation}
where spatial metrics are positive definite and the Newtonian gravitational 
constant is unity. Various properties were derived in Refs.\ 
\cite{Hayward:1997jp,sph,inq}, to which we refer for more detailed motivation 
of the definitions here. 

The Kodama vector \cite{Kodama:1979vn} is 
\begin{equation}
K=g^{-1}({*}dr)
\end{equation}
where $*$ is the Hodge operator in the space normal to the spheres of symmetry, 
i.e.\ 
\begin{equation}
K\cdot dr=0 \; \qquad\mbox{and}\qquad \quad g(K,K)=-g^{-1}(dr,dr) \;.
\end{equation}
Then $m$ is the Noether charge of the conserved energy-momentum density with 
respect to $K$. The Kodama vector gives a preferred flow of time, coinciding 
with the static Killing vector of standard black holes such as Schwarzschild 
and Reissner-Nordstr\"om. Note that $K$ is temporal, null or spatial on 
untrapped, marginal or trapped spheres respectively. 

The geometrical surface gravity was defined as \cite{Hayward:1997jp} 
\begin{equation}\label{kappa0}
\kappa=\frac{1}{2} \left({*}d{*}dr\right)\,,
\end{equation}
where $d$ is the exterior derivative in the normal space, or in terms of the 
metric $\gamma$ normal to the spheres of symmetry, 
\begin{equation}
\kappa = \frac{1}{2} \Box_\gamma r.
\end{equation}
Note also that $\kappa$ satisfies 
\begin{equation}
K^a\nabla_{[b}K_{a]}\cong\pm\kappa K_b\;,
\end{equation}
where $\cong$ denotes evaluation on a trapping horizon $r\cong2m$, similarly to 
the usual Killing identity. 

Then a trapping horizon is said to be {\em outer, degenerate} or {\em inner} if 
$\kappa>0$, $\kappa=0$ or $\kappa<0$ respectively. Examples are provided by 
Reissner-Nordstr\"om solutions. In vacuo, $\kappa$ is $m/r^2$, therefore 
reducing to the Newtonian surface gravity in the Newtonian limit, since $m$ 
reduces to the Newtonian mass \cite{Hayward:1997jp}. Thus it provides  a 
relativistic definition of the surface gravity of planets and stars as well as 
black holes. 

Any spherically symmetric metric can locally be written in dual-null 
coordinates $x^{\pm}$ as 
\begin{equation}\label{metric}
ds^2 = r^2 d\Omega^2-2e^{2\varphi}dx^+ dx^-
\end{equation}
where $d\Omega^2$ refers to the unit sphere and $(r,\varphi)$ are functions of 
$(x^+,x^-)$. There is still the freedom to rescale functionally $x^{\pm}\to 
\tilde{x}^{\pm} (x^{\pm})$. We wish  to use the generalized advanced 
Eddington-Finkelstein form 
\begin{equation}\label{EF}
ds^2 = r^2 d\Omega^2 + 2e^{\Psi}dvdr - e^{2\Psi} Cdv^2
\end{equation}
with $(C,\Psi)$ functions of $(r,v)$. Transforming from dual-null coordinates 
with $v = x^+$: 
\begin{equation}
dx^+ = dv \qquad \mbox{and}\qquad dx^- = \partial_v x^- dv +
\partial_r x^- dr\;,
\end{equation}
so one identifies 
\begin{equation}
e^{\Psi} = -e^{2\varphi} \partial_r x^- \quad \mbox{and}\quad
e^{2\Psi} C = 2e^{2\varphi}\partial_v x^- \;,
\end{equation}
which is possible if and only if $\partial_r x^- < 0$. Since we are assuming 
that $v$ is an advanced time, this means that any trapped ($C<0$) or marginal 
($C=0$) surface will be future trapped or marginal, as appropriate for black 
holes rather than white holes. 

Note that 
\begin{equation}
C = 1 - \frac{2m}{r}
\label{C}
\end{equation} is an
invariant, but $\Psi$ is not, due to the freedom $v \to \tilde{v}(v)$. We also 
have $K=e^{-\Psi}\partial_v$ and 
\begin{equation}
\kappa=\frac{1}{2} e^{-\Psi}\partial_r(e^{\Psi}C) = \frac{\partial_r C
  + C \partial_r \Psi}{2}\;,
\end{equation}
so 
\begin{equation}\label{kappa}
\kappa\cong\frac{\partial_r C}{2}\cong\frac{1-2\partial_r m}{2r}.
\end{equation}


\section{ Hamilton-Jacobi tunneling method}
The tunneling approach uses the fact that the WKB approximation of the 
tunneling probability along the classically forbidden trajectory from inside to 
outside the horizon has the form 
\begin{equation}
\Gamma \propto \exp\left(- 2\frac{\Im I}{\hbar}\right) \label{prob}
\end{equation}
where $\Im I$ is the imaginary part of the action $I$ on the classical 
trajectory, to leading order in $\hbar$, henceforth set to unity. If $\Im I$ is 
proportional to an energy parameter $\omega$, it takes a thermal form 
\begin{equation}\label{therm}
\Gamma\propto\exp\left(-\frac{\omega}{T}\right)
\end{equation}
which defines a temperature $T$. 

Consider a massless scalar field $\phi = \phi_0 \exp(iI)$ in the eikonal (or 
geometric optics) approximation, so that the amplitude $\phi_0$ is slowly 
varying and the action 
\begin{equation}\label{action}
I = \int\omega e^\Psi d v - \int kdr
\end{equation}
is rapidly varying. Here $e^{\Psi}$  is included to make $\omega$ and $I$ 
invariant, recalling the freedom $v \to \tilde{v}(v)$. Equivalently, 
$\omega=K\cdot dI=e^{-\Psi}\partial_vI$, $k = -\partial_r I$. Then the wave 
equation $\nabla^2\phi= 0$ yields the Hamilton-Jacobi equation 
\begin{equation}
g^{-1}(\nabla I,\nabla I)=0.
\end{equation}
In fact, the action (\ref{action}) and this equation are all that we need to 
assume here. It becomes simply 
\begin{equation}
2\omega k - Ck^2 = 0.
\end{equation}
Then $k = 0$ yields the ingoing modes, while $k = 2\omega/C$ yields the 
outgoing modes. Since $C\cong 0$ at a trapping horizon $r\cong r_0$, $I$ has a 
pole, which can be evaluated by $C \approx (r-r_0) \partial_r C$. Thus 
$k\approx\omega/\kappa(r - r_0 )$  if $\kappa\ncong 0$. Deforming the contour 
into the lower $\omega$ half-plane corresponds to deforming the contour into 
the upper $r$ half-plane, yielding an imaginary contribution 
\begin{equation}
\Im I \cong\frac{\pi \omega}{\kappa}.
\end{equation}
Then the particle production rate takes the thermal form (\ref{therm}) if the 
temperature is 
\begin{equation}
T\cong\frac{\kappa}{2\pi}.
\end{equation}
For this to be positive, $\kappa>0$, so the trapping horizon is of the outer 
type. Thus the method has derived a positive temperature if and only if there 
is a future outer trapping horizon. This remarkably confirms the local 
definition of black hole which was proposed previously on purely geometrical 
grounds \cite{Hayward:1993wb,Hayward:1997jp}. 

\section {Operational meaning: redshift}
Having derived the temperature $T$ formally, one may ask what it means 
operationally, i.e.\ in terms of what observers measure. First note that there 
is a preferred class of observers even in a non-static space-time, whose 
worldlines are the integral curves of $K$, who become static observers in the 
static case. These Kodama observers lie outside the horizon and have velocity 
vector $\hat K=K/\sqrt{C}$. Since $-I$ is the phase, the frequency measured by 
such observers is 
\begin{equation}
\hat\omega=\hat K\cdot dI=\frac{\omega}{\sqrt{C}}.
\end{equation}
Following the above method, such observers measure a thermal spectrum with 
temperature 
\begin{equation}
\hat T \approx \frac{T}{\sqrt{C}}
\end{equation}
to leading order near the horizon. More precisely, $\hat{T}\sqrt{C}\to T$ as 
$r\to2m$. The invariant redshift factor $\sqrt{C}$ (\ref{C}) is familiar from 
the Schwarzschild case \cite{BD}, where it reflects the acceleration required 
to keep an observer static. So this is the operational meaning of $T$: not that 
someone is measuring $T$ directly, but that the preferred observers just 
outside the horizon measure $T/\sqrt{C}$, which diverges at the horizon. Then 
$T$ itself can be interpreted as a redshift-renormalized temperature, which is 
finite at the horizon. 

We note that $\kappa$ (\ref{kappa}) is inequivalent to the Nielsen-Visser 
surface gravity \cite{NV}, which in these coordinates takes the form 
\begin{equation}
\tilde\kappa\cong \frac{1}{2r}(1-2\partial_rm-e^{-\Psi}\partial_vm) \;,
\end{equation}
though they coincide in the static case. Also, both are inequivalent to the 
Visser surface  gravity $e^\Psi\tilde\kappa$ \cite{Visser:2001kq}, which was 
derived as a temperature by essentially the same method as above, but in 
Painlev\'e-Gullstrand coordinates. The relative factor $e^\Psi$ is explained in 
the next section. The remaining difference can be traced to choice of time, 
since the action (\ref{action}) defines a frequency with respect to the time 
coordinate, and consequently different temperatures can be obtained. This 
reflects the generally unresolved issue of choice of time in quantum field 
theory on non-stationary space-times. One physical argument for choosing 
advanced time $v$ for a massless scalar field is that, in the classical limit, 
a particle follows null geodesics. A more thorough discussion will be given  in 
the longer article \cite{bht2}; here we merely stress the operational meaning 
of $T$. 


\section{ Static, asymptotically flat space-times}
The geometrical surface gravity $\kappa$ coincides with the usual definition of 
the Killing surface gravity $\kappa_\infty$ for standard examples of static 
black holes such as Schwarzschild and Reissner-Nordstr\"om. However, it does 
not coincide if $\Psi\not\cong0$, requiring further explanation. 

Note first that the unit normalization of the Killing vector 
$K_\infty=\partial_t$ at infinity is crucial to the definition of 
$\kappa_\infty$, since otherwise $K_\infty$ can be rescaled by a positive 
constant and $\kappa_\infty$ scales likewise. This scaling, however, cannot be 
known locally. 
This problem is illustrated by two Schwarzschild regions matched across a 
static shell outside the horizon, such that $\Psi\not\cong0$: if surface 
gravity is to be a local quantity, one would expect to define it with the 
normalization appropriate to the interior region containing the horizon, rather 
than the normalization at infinity. 

Consider static metrics in the form 
\begin{equation}
ds^2 =r^2 d\Omega^2
+C^{-1}dr^2  -Ce^{2\Psi} dt^2 \label{vw}
\end{equation}
where $(C,\Psi)$ are henceforth functions of $r$ alone, the notation being 
consistent with the metric (\ref{EF}). Then $\kappa_\infty$ is defined by 
\begin{equation}
K_\infty^a\nabla_bK_{\infty a}\cong\kappa_{\infty}K_{\infty b}
\end{equation}
and yields 
\begin{equation}
\kappa_{\infty}\cong e^{\Psi}\kappa.
\end{equation}
This relative factor stems from our use of the Kodama vector $K$ instead of the 
static Killing vector 
$K_\infty=e^\Psi K$, 
since the latter does not exist in dynamic cases. Thus, we can deal in a 
unified way with such situations as an accreting black hole which settles down 
to a static state, or a static black hole which starts to evaporate. In our 
opinion, the relative  factor $e^\Psi$ can be explained as follows, first 
noting that a textbook method derives the gravitational redshift of light along 
a given ray \cite{Wald}: 
$\sqrt{-g(\partial_t,\partial_t)}\hat\omega=e^\Psi\sqrt{C}\hat\omega$ is 
constant along the ray. 

If the space-time is asymptotically flat, with $(t,r)$ being Minkowski 
coordinates as $r\to\infty$, then $C\to1$, $\Psi\to0$ and $\partial_t\to K$. 
Note that it is precisely here where the generally non-invariant $\Psi$ 
acquires a specific meaning. Then the frequency measured by static observers at 
infinity is 
\begin{equation}
\omega_\infty=e^\Psi\sqrt{C}\,\hat\omega
\end{equation}
and the corresponding temperature measured by such observers is 
\begin{equation}
T_\infty=e^\Psi\sqrt{C}\,\hat T,
\end{equation}
which is the famous Tolman relation \cite{Tol}. Thus 
\begin{equation}
T_\infty\cong e^\Psi T,
\end{equation}
which indeed corresponds to $\kappa_\infty\cong2\pi T_\infty$. Such 
considerations suggest to interpret $e^{\Psi}$, which in general measures the 
discrepancy between the Killing and the Kodama temperatures, as an 
interpolating factor between $T_\infty$ and $T$, appearing as a relative 
redshift between the horizon and infinity, or as a gravitational dressing 
effect, since $e^{\Psi}$ enhances the redshift over the spatial curvature. 


These results seem to suggest that the appropriate local temperature at the 
horizon is $T$ and generally not $T_\infty$ even in the static case. Likewise, 
the local surface gravity is $\kappa$ and generally not the textbook definition 
$\kappa_{\infty}$. Recall that the physical interpretation of $\kappa_\infty$ 
is the force at infinity per unit mass required to suspend an object from a 
massless rope just outside the horizon \cite{Wald}. This ``surface gravity at 
infinity'' would seem to be an oxymoron. Certainly this is not how Newtonian 
surface gravity is defined, as the local gravitational acceleration. Recall as  
above that $\kappa$ reduces to the latter in vacuo.

\section{ Extremal limit} As an example, consider the charged stringy black 
hole, which represents a non-vacuum solution of Einstein-Maxwell dilaton 
gravity in the string frame \cite{gibbons,garf}: 
\begin{equation} ds^2=r^2d \Omega^2
+\frac{dr^2}{\left(1-a/r\right) \left(1-b/r\right)}
-\left( \frac{1-a/r}{1-b/r}\right)dt^2
\end{equation}
where $a>b>0$. The horizon radius is $r\cong a$. 

For this example, the extremal limit as defined by global structure is $b 
\rightarrow a $. However, the Killing surface gravity 
\begin{equation}
\kappa_{\infty}\cong\frac{1}{2a}
\end{equation}
does not vanish in this limit. Garfinkle et al.\ \cite{garf} noted this as 
puzzling, since extremal black holes are expected to be zero-temperature 
objects. 

Remarkably, the geometrical surface gravity \eref{kappa0} 
\begin{equation}
\kappa\cong\frac{a-b}{2 a^2}
\end{equation}
vanishes in the extremal limit. Thus the gravitational dressing effect lowers 
the temperature to its theoretically expected value. 

We conjecture that this is  true in general. Indeed, past experience with 
extremal black holes showed that the horizon of these objects is not only a 
zero but also a minimum of the expansion $\theta_+=\partial_+A/A$ of the 
radially outgoing null geodesics, $\theta_+$ becoming positive again on 
crossing the horizon. Thus $\partial_-\theta_+\cong0$ should be the appropriate 
definition of an extremal black hole. Since 
$\kappa=-e^{-2\varphi}\partial_-\partial_+r$, this is equivalent to 
$\kappa\cong0$. 

\section{Concluding  Remarks}

We conclude that dynamical black holes do indeed possess a local temperature 
$T$, with the operational meaning that it determines the redshifted temperature 
$T/\sqrt{1-2m/r}$ measured by Kodama observers just outside a trapping horizon. 
Moreover, the method works precisely for future outer trapping horizons, as 
proposed previously to define black holes on purely geometrical grounds, and 
$T=\kappa/2\pi$ in terms of the geometrically defined surface gravity $\kappa$. 
This confirms the quasi-stationary picture of black-hole evaporation in the 
early stages. 

The derivation holds formally even in regimes where one normally expects a 
semi-classical approximation to break down. With this qualification, it 
strongly suggests that evaporation proceeds until $\kappa\to0$. While this is 
reminiscent of quasi-stationary arguments, it has a different meaning, since 
$\kappa$ is generally not the surface gravity of a static black hole with the 
same mass, charge or whatever other parameters in a given model. It also 
encodes information about the dynamic space-time geometry, such as the rate of 
evaporation. This may be of relevance to the information puzzle. 

A common idea is that evaporation results in an extremal remnant 
\cite{Giddings:1995, Diba:2002hb}.  For instance, an outer ($\kappa>0$) and 
inner ($\kappa<0$) trapping horizon might asymptote to the same null 
hypersurface, effectively forming a degenerate ($\kappa=0$) trapping horizon. 
Another idea is that the outer and inner trapping horizons merge smoothly at a 
single moment of extremality where $\kappa$ vanishes \cite{02bh20}. The results 
here are consistent with either picture. 

Finally, we note that a minor modification derives a positive temperature for 
past inner trapping horizons, namely to use a retarded time $u$ instead of $v$, 
as will be discussed in more detail elsewhere. For future inner or past outer 
trapping horizons, there is formally a negative temperature, but the physical 
meaning is debatable. 

\ack 

SAH thanks Ted Jacobsen and Alex Nielsen for discussions. SAH was supported by 
the National Natural Science Foundation of China under grants 10375081, 
10473007 and 10771140, by Shanghai Municipal Education Commission under grant 
06DZ111, and by Shanghai Normal University under grant PL609. 

\noindent RDC wishes to thank the INFN - Gruppo Collegato di Trento and the 
Department of Physics at the University of Trento, where part of this work has 
been done.

\end{document}